# Об индексах цитирования научных работ
## А.С. Холодов[1,2,3]

[1] Институт автоматизации проектирования РАН,
Россия, 123056, г. Москва, 2-я Брестская ул., д. 19/18
[2] Московский физико-технический институт,
Россия, 141700 г. Долгопрудный, М. о., Институтский пер., 9
[3] Балтийский федеральный университет им. Им. Канта,
Россия, 123056, г. Калининград, ул. А. Невского, 14
E-mail: xolod@crec.mipt.ru

Рассмотрены некоторые вопросы формирования наукометрических параметров. Предложены два новых индекса цитирования научных работ, более полно и более объективно оценивающих значимость научных работ авторов и авторских коллективов.

Ключевые слова: наукометрия, индекс цитирования научных работ.

Several questions of scientometrics parameters organization are considered. Two new indices for scientific works citation analysis are proposed. They provide more detailed and reliable scientific significance assessment of individual authors and scientific groups basing on the publication activity.

## 1. Введение

С появлением современных информационных технологий и мощных поисковых систем огромные объемы цифровых версий накопленных ранее и вновь создаваемых научных знаний стали гораздо более доступны широкому кругу пользователей и, в том числе, вызвали новую волну интереса к проблемам наукометрии, связанную, с их систематизацией ([1] и др.). Активизировались попытки количественных оценок значимости отдельных работ, рейтинговых оценок их авторов и авторских коллективов. Если ранее автор научных работ вел свой личный список печатных (иногда и рукописных) работ (известная форма 4), заверенная ученым секретарем организации, копия которого использовалась не так уж часто (при приеме на научную должность, при защите диссертаций и т.п.), сейчас гораздо более объемная информация (включая числа цитирований из разных поисковых систем: WoS, Scopus, e-library, Google Scholar и др., всевозможные индексы цитирования и импакт-факторы) все более активно стали использоваться работодателями, в том числе для оценки эффективности работы научных сотрудников. В определенной степени число цитирований отражает значимость работы, но абсолютизировать этот критерий нет достаточных оснований. В литературе приводится достаточно много веских оснований и примеров на эту тему. В связи с этим возникает проблема полноты и достоверности подобной информации, а также объективности различных используемых критериев, без которых ценность подобных систем даже не нулевая, а сильно отрицательная. На несовершенство существующих систем только количественной оценки научной деятельности указывалось во многих публикациях (см., например, [2-12] и цитируемая в них литература), а также в решениях регулирующих научную деятельность органов ряда стран [13], на что указано, например, в докладе А.Н.Паршина 10.09.2013 на заседании Отделения математических наук РАН [14]. Тем не менее, процесс внедрения подобных систем в нашей стране продолжается ускоряющимися темпами и существует необходимость хотя бы сгладить отрицательные его последствия.

Одной из таких задач является усиление возможностей контроля авторов с помощью удобного инструментария за действиями автоматизированных систем (точнее их разработчиков и коммерческих структур, работающих в этом виде бизнеса). Положительным примером может служить система работы с индивидуальными профилями авторов Google Scholar, в которой, в отличие от аналогичной системы WoS, для авторов помимо функций добавить, удалить и экспортировать есть функция объединения ссылок, позволяющая исправлять многочисленные ошибки перевода и ссылок в списках литературы цитирующих работ. Хотя и в этой системе необходимо внести дополнения, предоставляющие авторам возможность подсказывать системе пропущенные ее цитирования (соответствующей ссылкой на цитируемую и цитирующую работы). Это стимулировало бы разработчиков систем цитирования расширять используемые базы данных и позволило бы существенно повысить достоверность поисковых результатов. В большинстве других систем вмешательство авторов не допускается (Scopus и др.) либо крайне затруднено (e-library и др.).

Необходимо также полностью исключить учет такого чисто издательского параметра как импакт-фактор журнала, никак не влияющего на научную значимость публикуемой в нем работы, а, наоборот, завышающего значимость слабоцитируемых работ из журнала с высоким импакт-фактором за счет высокоцитируемых. В частности, в работе [13] приведены британские методы оценки научной деятельности, а в решении Комитета по науке и технологиям Палаты общин британского парламента (2004г.) изложено требование не использовать импакт-фактор журналов для оценки опубликованных в них работ. Соответствующие (и немалые) преференции публикующийся в престижном журнале автор и без того получает за счет более широкой аудитории читателей его статьи. Аналогичная дискриминация возникает между публикациями в англоязычных и национальных журналах. Здесь также определенное преимущество за счет более широкой аудитории имеют англоязычные публикации. Лишь частично это преимущество нивелируется для публикаций в национальных журналах, имеющих перевод на английский язык.

Не менее важен и контроль над некоторыми авторами, злоупотребляющими представившимися возможностями манипулировать цитированиями собственных публикаций (организация перекрестных цитирований, не соответствующие содержанию публикуемой статьи ссылки на собственные работы и др.). Следует, например, принять за правило не употреблять (или не учитывать в системах цитирования) более 10-15% от общего списка литературы ссылок на собственные работы автора. Это лишь часть имеющихся недоработок в наиболее распространенных цитирующих системах, требующих соответствующих договоренностей в научном сообществе.

Нельзя не обращать внимания на то, что показатели одного и того же автора в различных существующих системах могут отличаться даже не в разы, а в десятки раз, на что неоднократно указывалось в литературе (например [15] и др.). В качестве примера в таблице 1 приведены данные автора данной работы, заимствованные из разных претендующих на универсальность систем цитирования (данные на 15.01.2014г., [16]). Думаю, что подобный разброс не только у меня и является не исключением, а весьма распространенным явлением. Печально, что, несмотря на очевидные прорехи, использование подобных низкокачественных систем, в частности, в нашей стране активно продавливается на весьма высоком уровне. В таких условиях необходимо предоставить авторам выбор устраивающей его системы цитирования во всех затрагивающих его интересы случаях, что неизбежно приведет к большей унификации различных систем цитирования и упорядочиванию существующих в данной области проблем.

Таблица 1

| No 1.4 | Общее число работ | Число работ, процитированных хотя бы один раз | Общее число цитирований | Среднее число цитирований одной работы | Среднее число цитирований одной цитировавшейся работы | Индекс Дж.Э. Хирша | Индекс $I_{10}$ |
|---|---|---|---|---|---|---|---|
| Google | 200 | 68 | 737 | 3.7 | 10.8 | 14 | 20 |

| Scholar | | | | | | | |
|---|---|---|---|---|---|---|---|
| e-library | 66 | 54 | 417 | 6.32 | 7.72 | 8 | 7 |
| РИНЦ | 28 | - | 42 | 1.5 | - | - | - |
| Scopus | 32 | - | 16 | 0.5 | - | 3 | - |
| WoS | 16 | 12 | 23 | 1.4 | 1.92 | 2 | 1 |

**2. Сравнение некоторых индексов цитирования и формирование индекса на базе системы основных параметров функции цитирования**

Появившиеся в последние годы разнообразные индексы цитирования научных работ базируются на различных параметрах и свойствах функции цитирования $C(r)$ – зависимости числа цитирований $C_r$ работ с номерами $r$ (расположенных в порядке уменьшения $C_r$ от $C_{max} = C_1 = C(1)$ до $C_{min} = C(R+1) = 0$). Здесь принято, что в промежутках между целочисленными номерами $r$ используется линейное восполнение функции $C(r)$ по целочисленным ее значениям $C_r$. Сразу следует оговориться, что работы авторов с номерами $(R+1, ..., R_0)$ не имеющие ни одного цитирования, целесообразно никак не учитывать, чтобы можно было работать с содержательной частью функции $C(r)$ и не поощрять любителей публиковать большое число, но малозначимых работ. Здесь $R_0$ общее число работ автора, $R$ – число работ, цитировавшихся к рассматриваемому моменту времени хотя бы один раз. Достойные работы со временем (весьма разным для разных областей знаний) пополняют список цитируемых работ автора, а некоторые (чаще всего это тезисы докладов на конференциях) не удостаиваются цитирования даже ее автором. По приведенному определению $C(r)$ монотонно убывающая функция на отрезке [1,R]. Примеры таких функций приведены на рис. 1-4 ломаными линиями и демонстрируют, что их объективное сопоставление друг с другом далеко не тривиальная задача, как это предлагается во всех существующих системах и индексах цитирования. Для монотонно убывающей функции действительного аргумента $C(r)$ помимо минимального и максимального значений в области ее определения весьма значимыми и легко вычисляемыми являются интеграл $C_\Sigma = \sum_{r=1}^{R} C_r$ и ее среднее значение $C_s = C_\Sigma / R$, играющие значительную роль при использовании различных функций статистических распределений.

Некоторые из определяющих поведение функции $C(r)$ параметров $R_0, R, C_\Sigma, C_{max}, C_s$, поврозь используются в качестве индексов цитирования в существующих системах. Кроме того в качестве индексов цитирования используются: число цитирований 10 наиболее цитируемых работ автора $C_{10}$ (использовалось, например, в конкурсном отборе ведущих ученых [17]); число работ, процитированных не менее 10 раз ($I_{10}$ - индекс, используется в системе цитирования Google Scholar [18]); число авторов, общее число цитирований $C_\Sigma$ у которых более 1000 [10]; индекс Дж.Э. Хирша (h-индекс) [19]; его некоторые модификации - индекс Л.Эгга (g-индекс, [20]), m-индекс $m = h/n$ (учитывающий разность n между текущим возрастом автора и его возрастом в момент первой публикации, [19]) и, наверное, многие другие. Возникновение столь большого числа индексов цитирования связано с попытками определить рейтинг ученого или авторских коллективов посредством одного количественного параметра, что, как отмечалось, является довольно сложной задачей. Наиболее популярным сейчас является h-индекс: точка пересечения с функцией $C(r)$ прямой $C = r$, исходящей под углом $45°$ к оси абсцисс из начала координат ($r = 0, C = 0$), т.е. точка, в которой номер работы $r$ и число ее цитирований $C_r$ связаны неравенством $r \leq C_r$, а для следующей работы $r+1 > C_{r+1}$. В индексе Л.Эгга для усиления влияния наиболее цитируемых работ прямая $C = r$ заменяется параболой $C = r^2$. Естественно возникает вопрос: почему $45°$ а не $0°$ (т.е. число цитируемых работ $R$) или не $90°$ (т.е. $C_{max}$)? Почему $C = r^2$, а не $C = r^3$, почему $I_{10}$, а не $I_{100}$ и т.д. Более естественным представляется попытаться свести все значимые параметры функции $C(r)$ воедино, например, $R$, $C_\Sigma$, $C_s$ непосредственно и $C_{max}$ косвенно. Для этого в алгоритме Дж.Э.Хирша вместо прямой $C = r$ используем прямую $C = C_s r = C_\Sigma r / R$, тем самым косвенно учитывая и значение $C_{max}$. Назовем этот индекс $Kh_1$. По сложности вычисления он практически не отличается от индекса Дж.Э.Хирша, а качество представления функции $C(r)$ (оценку распределения цитирований) должен обеспечивать гораздо лучше, т.к. в той или иной степени учитывает практически все значимые параметры распределения.

Игнорирование почти во всех системах $C_{max}$ представляется ошибочным. Частично, хотя и довольно слабо, он учитывается в $C_{10}$, еще слабее он влияет на $C_s$. В качестве примера из вычислительной механики можно привести наиболее цитируемую работу с $C_{max} = 2589$ автора распределения 1.1 из таблицы 2 - выдающегося ученого, одного из основоположников современной вычислительной математики, в начале пятидесятых годов создавшего уникальный численный метод для решения нелинейных систем уравнений гиперболического типа (уравнений Эйлера) и положившего начало активно развивающемуся и сейчас научному направлению в этой области (рис.1, таб.2). Между тем эта работа, с описанием созданного в самом начале 50-х годов метода, поступившая по ряду причин в редакцию одного из русскоязычных журналов только в 1956г., была опубликована лишь в 1959г., да и то в периодическом сборнике. Две примерно равноценные линеаризованные версии этого метода (менее ресурсоемкие по сравнению с оригиналом, но, также как и оригинал, требующие создания новой модификации метода для каждой новой нелинейной системы уравнений гиперболического типа), разработанные другими авторами и опубликованные в восьмидесятых годах в англоязычных журналах разных стран имеют существенно различающиеся числа цитирований (одна - более шести тысяч цитирований, а другая – примерно 100). Более ранняя, более универсальная и менее ресурсоемкая по сравнению и с оригиналом и с упомянутой ее более поздней версией линеаризация автора данной работы, опубликованная в русскоязчном журнале (имеющим и англоязычный перевод) в конце семидесятых годов, имеет 22 цитирования. К сожалению, оставляет желать лучшего (и со временем все более ухудшается) культура научных цитирований. Вот лишь еще один пример на эту тему. В работе [21] один из разделов носит название «Аппроксимация Годунова и главные результаты», однако среди 28 ссылок в списке литературы этой статьи нет ни одной ссылки на работы автора этого метода С.К.Годунова. Имена многих авторов действительно основополагающих научных результатов со временем становятся нарицательными и вместо их работ упоминаются лишь имена (законы Архимеда, Ньютона, уравнения Эйлера, теория относительности Эйнштейна, проблемы Гильберта, теорема Годунова и т.д.). Но тогда может быть стоит и этот вид цитирования включать в число авторских цитирований? Технически это не сложно (вопрос только в том, к каким работам автора относить цитирования такого рода, но это лишь вопрос договоренностей в научном сообществе. Такое расширение понятия цитирований было бы более объективной характеристикой вклада автора в науку, т.к. сейчас, например, по запросу «Isaac Newton», а тем более «Исаак Ньютон» любая из систем цитирования выдаст отнюдь не рекордные числа. Таких примеров, подтверждающих необходимость совершенствования систем цитирования научных работ, можно приводить бесконечно много и в самых разных областях. Существенно то, что лишь одной упомянутой выше работой автор распределения 1.1 заслуживает того, чтобы его индекс цитирования был существенно выше, чем у его соседей – авторов распределений 1.2 и 1.3 (также очень достойных ученых, но все-таки уровнем пониже, а по индексу Дж.Э.Хирша они практически не различимы). Именно это и демонстрирует предлагаемый индекс $Kh_1$. Т.к. автору данной работы (номер распределения 1.4) не известна возможная реакция авторов других распределений на приводимые в данной работе сопоставления, они обезличены, хотя это данные из вполне конкретных профилей открытого доступа Google Scholar сотрудников одного из подразделений российского научно-образовательного учреждения (кроме автора функции цитирования 1.1).

Вместе с тем следует иметь ввиду, что при увеличении возможности авторов влиять на указанное в собственном профиле число процитированных работ, например, уменьшая число малоцитируемых, повышается $C_s$ (в пределе до $C_{max}$ при $R=1$), а это приводит к соответствующему росту индекса $Kh_1$. Для того, чтобы исключить при необходимости влияние параметра $R$, можно использовать индекс $Kh_2 = \sqrt{C_\Sigma}$, который для заданного общего числа цитирований $C_\Sigma$ является максимально допустимым значением индекса Дж.Э.Хирша (если бы $C_\Sigma$ состояло из $R = Kh_2$ работ, у каждой из которых было бы одинаковое и равное $Kh_2$ число цитирований). Этот экзотический пример свидетельствует о том, что ставший наиболее популярным *h*-индекс есть только мера отклонения реального распределения $C(r)$ от сформулированного автором данного индекса идеала. Но вот идеален ли идеал - это большой вопрос. Как неоднократно отмечалось в литературе, одним из наиболее существенных недостатков индекса Дж.Э.Хирша является то, что при его использовании не различаются авторы, имеющие $R$ процитированных по одному разу работ и авторы, имеющие 1 процитированную $R$ раз работу. Их *h*-индекс равен единице, что в $R$ раз меньше предельного для описанного выше специального распределения с $C_\Sigma = R^2$. В описанном выше примере, если бы автор распределения 1.1 имел только 1 работу с $C_{max} = C_\Sigma = 2589$ и имел *h*-индекс равный 1, трудно поверить, что его рейтинг почти в 51 раз меньше, чем у автора работ, имевшего бы 51 работу, каждая из которых была бы процитирована по 51 раз. Опасения, связанные с чрезмерным влиянием $C_{max}$ при использовании $Kh_1$ представляются несостоятельными, т.к. экзотические распределения, состоящие лишь из одной высокоцитируемой работы вряд ли возможны (если не создавать подобные ситуации специальным манипулированием). Для молодых ученых значения $Kh_1 = C_{max}$, вполне допустимы из-за не очень больших $C_{max}$.

Индекс $Kh_1$ всегда больше *h*-индекса и совпадает с ним только в предельном случае цитируемых по одному разу всех $R$ работ. Другое предельное значение индекса $Kh_1$ равно $C_{max}$ при $R=1$. Если ввести индекс $Kh_3$, определяемый аналогично $Kh_1$, но с использованием прямой $C = r\sqrt{C_\Sigma}$, можно совместить максимально возможные значения индексов $Kh_3 = h = R$ и *h*-индекса при заданном $C_\Sigma = R^2$ и одинаковых значениях числа цитирований каждой из $R$ работ $C(r) = C_{max} = R = const$. Эта версия индекса (как и индексы Дж.Э.Хирша и Л.Эгга) позволяет исключить влияние параметра $R$, допускающего различные манипуляции в системах цитирования с доступом автора к списку его цитируемых работ, что, как отмечалось выше, объективно необходимо для большей достоверности функции цитирования $C(r)$.

При любом максимально объективном выборе индекса цитирования, он должен быть лишь дополнением системы экспертной оценки значимости работ автора. Причем дополнением обязательным, т.к. случаи необъективных, в том числе ангажированных экспертных оценок, не так уж редки. В приведенном выше примере с функцией цитирования 1.1, разумеется, были экспертные оценки рецензентов. Но, скорее всего, тогдашние ортодоксальные математики не очень признавали за значимый результат, полученный с использованием только нарождавшейся вычислительной математики, а столь же ортодоксальные механики также не очень доверяли численным расчетам. Автор данной публикации на заре своей научной деятельности еще был свидетелем подобных баталий.

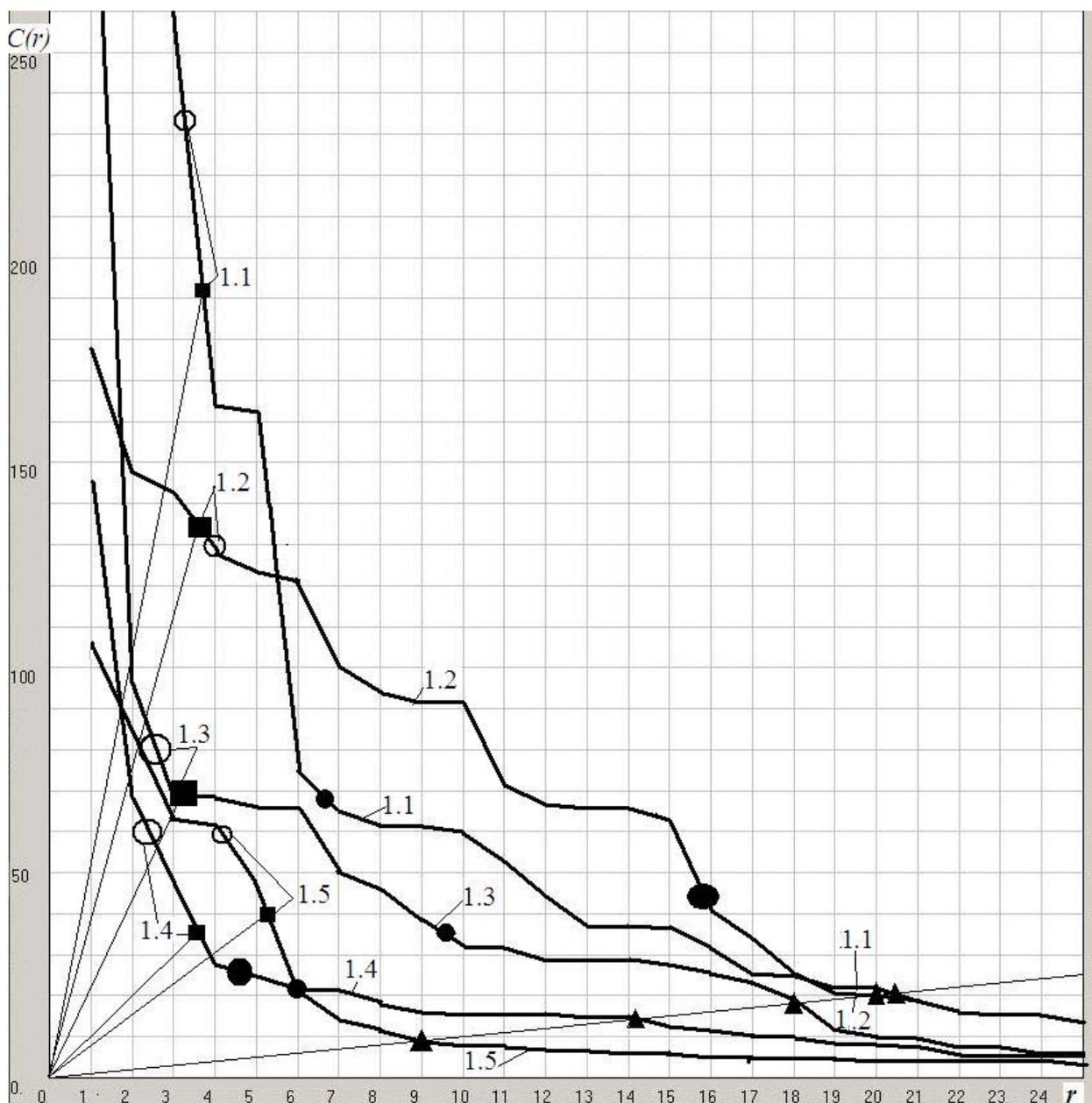

Рис.1. Индекс Дж.Э.Хирша – треугольники; индекс $Kh_1$ – квадраты; индекс $Kh_2$ – темные кружки; индекс $Kh_3$ – светлые кружки.

### 3. Результаты вычислительных экспериментов

Для сопоставления различных индексов цитирования были использованы обезличенные данные конкретных авторских профилей открытого доступа Google Scholar (как наиболее информативной и наименее ангажированной системы) на 18.06.2013г. [16]. Некоторые результаты этих сопоставлений представлены на рис. 1-3 и в таб. 2-4. Как отмечалось, использованные данные относятся к коллективу из 30 авторов (с номерами функций цитирования 1.1-1.5, 2.1-2.13, 3.1-3.12, столбцы 1 таб.2-4) в основном из одного подразделения российского научно-образовательного учреждения, имеющих возраст от 27 до 84 лет (средний возраст в каждой из трех подгрупп, соответственно: 63, 62 и 39 лет) и столь же разные наукометрические показатели: по общему числу работ от 1 до 239 (столбцы 2), по числу цитируемых работ от 1 до 142 (столбцы 3), по общему числу цитирований $C_\Sigma$ от 2 до 4559 (столбцы 4), по максимальному числу цитирований одной работы от 2 до 2589 (столбцы 6), по среднему числу цитирований одной работы $C_s$ от 1.8 до 51.8 (столбцы 7). Нумерация функций цитирования проведена в порядке убывания $C_{max}$. Данная выборка хотя и не очень большая количественно, но вполне представительная для направления, связанного с математическим моделированием достаточно широкого круга задач: теория численных методов; вычислительная механика, включая аэрогазодинамику, механику деформируемого твердого тела и сейсмодинамику, медицинскую макромеханику и биологическую микромеханику; вычислительная физика, включая физику технической и ионосферно-магнитосферной плазмы, астрофизику; функционирование различных сетевых систем, включая вычислительные модели различных сетей человеческого организма, транспортных, электроэнергетических, информационных сетей и др. Поскольку сравниваемые параметры изменяются в весьма широких диапазонах, что довольно сложно представить на едином рисунке, было сформировано три подгруппы из условно «высокоцитируемых» авторов с $C_{max}$ от 106 до 2589, «среднецитируемых» авторов с $C_{max}$ от 12 до 97 и «малоцитируемых» авторов с $C_{max}$ от 2 до 13, составленную в основном из молодых научных сотрудников. В таблицах 2-4 функции цитирования расположены по убыванию $C_{max}$. В столбцах 5, 8-14 таб. 2-4 представлены соответственно: индекс $C_{10}$, $h$-индекс Дж.Э.Хирша, $g$-индекс Л.Эгга, $m$-индекс, индекс $I_{10}$, индекс $Kh_1$, индекс $Kh_2$ и индекс $Kh_3$. Видно, что при выборе вместо $C_{max}$ другого параметра, их взаимное расположение внутри таблиц изменилось бы довольно существенно. На рисунках 1-3 функции цитирования $C(r)$ представлены ломаными линиями с указанием номера функции цитирования автора. Индексы Дж.Э.Хирша $h$ показаны как точки пересечения тонкой прямой $C = r$ с функциями цитирования $C(r)$ треугольными точками, индексы $Kh_1$ - как точки пересечения прямых $C = C_s r$ с функциями цитирования $C(r)$ квадратными точками, индексы $Kh_2$ - темными круглыми точками, а индексы $Kh_3$ - светлыми круглыми точками на соответствующих кривых $C(r)$. Для некоторых функций цитирования (при относительно малых $C_{max}$ и больших $C_\Sigma$) индекс $Kh_2$ может принимать большие, чем $C_{max}$ значения.

Из представленных на рис.1 данных видно, что наиболее цитируемые авторы сопоставляемой выборки (с номерами 1.1-1.3) практически не различаются по $h$-индексу, что, как и данные по индексу Л.Эгга, явно не соответствует реальному их рейтингу (и по количественным характеристикам – параметрам $C_\Sigma, C_s, C_{max}$, индексу $C_{10}$ и по экспертной оценке автора данной публикации, достаточно хорошо знающего работы всех сопоставляемых авторов. Индексы $Kh_1$, $Kh_2$ и $Kh_3$, включающие интегральные характеристики функций цитирования, более объективно отражают реальный рейтинг группы авторов, сопоставляемых на рис.1.

Для аналогичных данных на рис.2, как и выше, индекс $Kh_1$ позволяет более объективно детализировать рейтинг авторов, имеющих близкие значения индекса Дж.Э.Хирша (с функциями цитирования 2.1-2.3, 2.9 и 2.5, 2.7, 2.11, 2.13). Несколько выпадают из общей картины данные автора 2.9, имеющего самое большое в этой подгруппе число цитирований $C_\Sigma$, но имеющего сравнительно небольшое $C_s$ из-за небольшого $C_{max}$, а также из-за самого большого списка цитируемых работ $R$ и, как следствие, относительно низкие индексы $Kh_1$, $Kh_3$ (при очень высоком индексе $Kh_2$). Аналогично, относительно низкие в сравнении с их непосредственными соседями значения $C_s$ авторов 2.4, 2.5, 2.8 и 2.10 привели к некоторому (в сравнении с соседями) снижению их индексов $Kh_1$. Отмеченные исключения являются следствием влияния параметра $R$ и отсутствуют у индексов $Kh_2$, $Kh_3$ не использующих этот параметр.

В третьей подгруппе (рис.3, авторы с функциями цитирования 3.1-3.12) следует отметить полную непригодность для малоцитируемых авторов индексов цитирования $I_{10}$ и Л.Эгга, а *m*-индекса – для всех возрастных категорий (для чего достаточно сопоставить *m*-индекс функции цитирования 1.1 с его значениями у остальных распределений). Индекс Дж.Э.Хирша для первых четырех из них (3.1-3.3, 3.5) изменяется в пределах 4-5, а для всех остальных равен 1- 2, что, конечно не отражает ни текущий их рейтинг, ни их потенциальные возможности. Использование здесь индексов $Kh_1$, $Kh_3$ позволяет, как и в первых двух подгруппах, детализировать их текущий рейтинг (что наиболее важно именно для молодых ученых), а также более объективно прогнозировать динамику их роста.

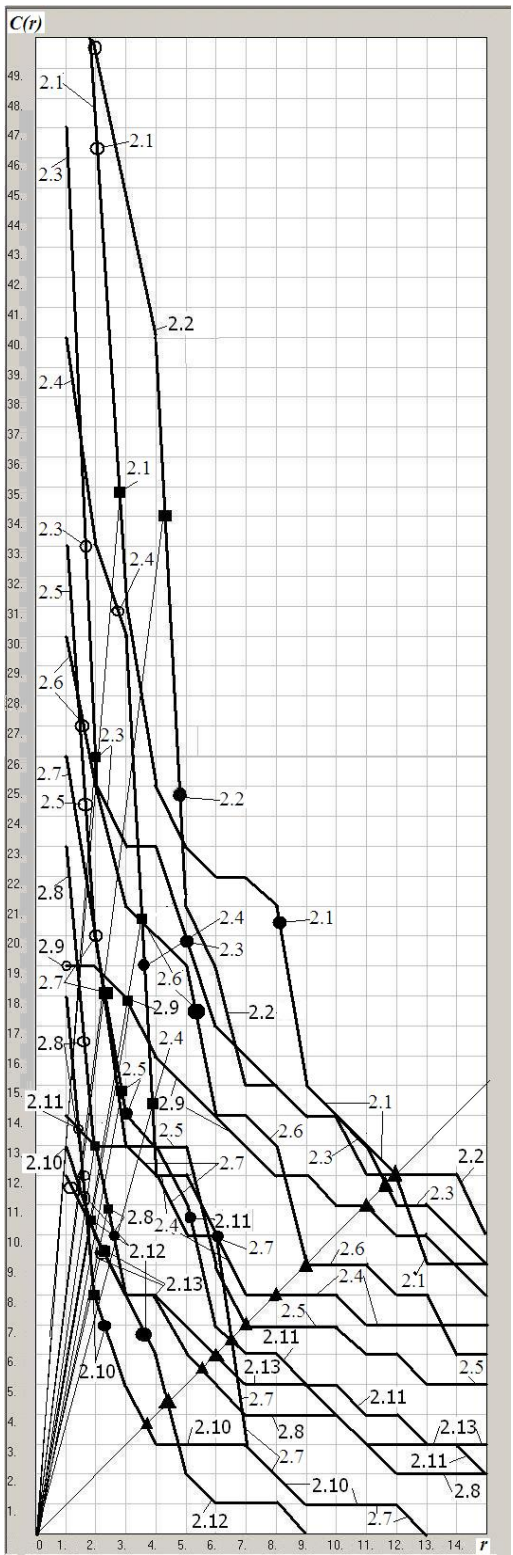

Рис.2. Индекс Дж.Э.Хирша – треугольники; индекс $Kh_1$ – квадраты; индекс $Kh_2$ – темные кружки; индекс $Kh_3$ – светлые кружки.

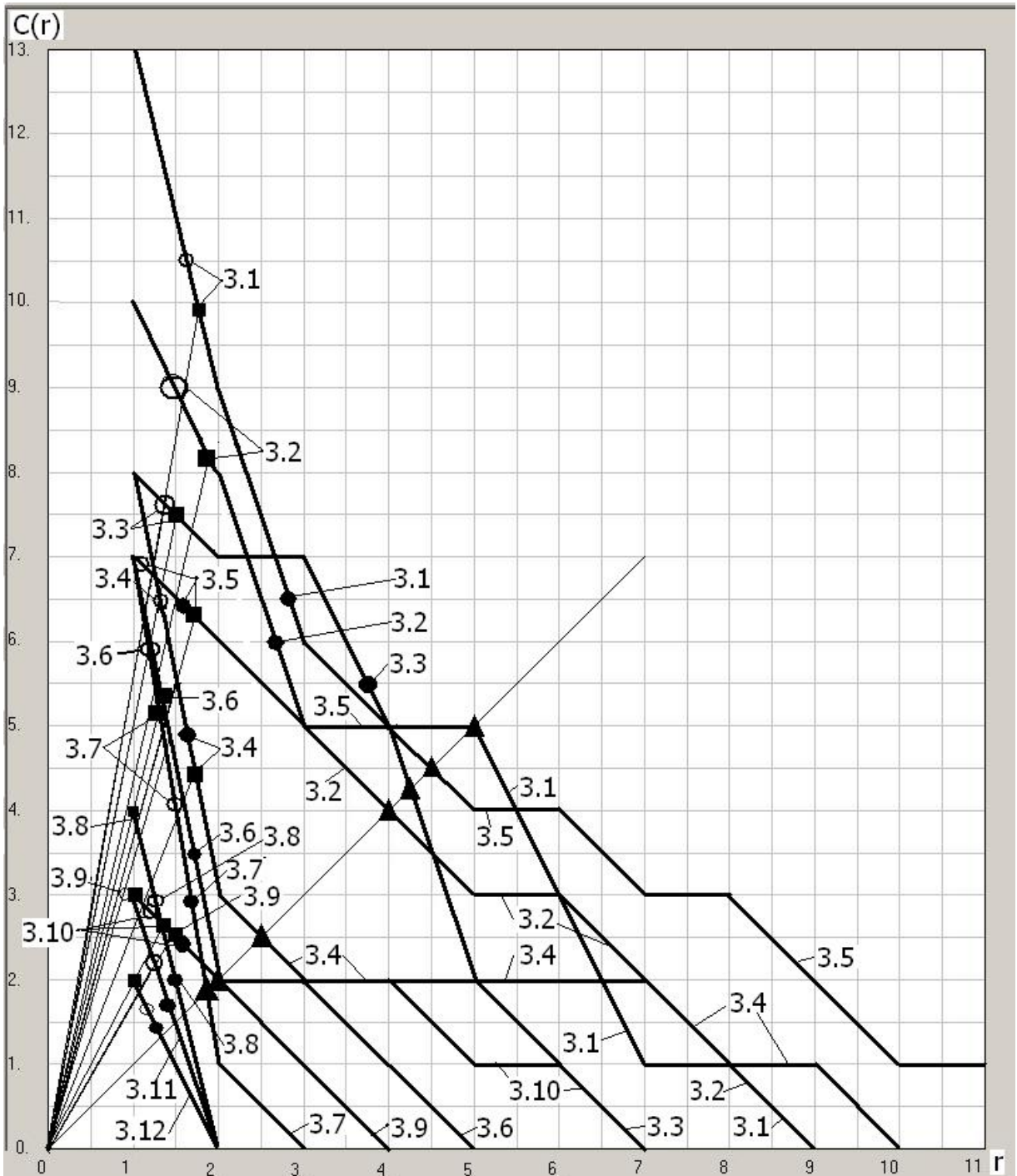

Рис.3. Индекс Дж.Э.Хирша – треугольники; индекс $Kh_1$ – квадраты; индекс $Kh_2$ – темные кружки; индекс $Kh_3$ – светлые кружки.

Таблица 2

| No | $R_0$ | $R$ | $C_\Sigma$ | $C_{10}$ | $C_{max}$ | $C_s$ | $h$ | $g$ | $m$ | $I_{10}$ | $Kh_1$ | $Kh_2$ | $Kh_3$ |
|---|---|---|---|---|---|---|---|---|---|---|---|---|---|
| 1 | 2 | 3 | 4 | 5 | 6 | 7 | 8 | 9 | 10 | 11 | 12 | 13 | 14 |
| 1.1 | 239 | 88 | 4559 | 3853 | 2589 | 51.8 | 20 | 49 | 0.3 | 35 | 191 | 67.5 | 225 |
| 1.2 | 105 | 51 | 1875 | 1220 | 178 | 36.8 | 20 | 81 | 0.7 | 27 | 133 | 43.3 | 130 |
| 1.3 | 77 | 59 | 1218 | 855 | 318 | 20.6 | 18 | 49 | 0.8 | 21 | 69.0 | 34.9 | 80 |
| 1.4 | 198 | 67 | 682 | 407 | 145 | 10.2 | 14 | 16 | 0.3 | 18 | 36.2 | 26.1 | 60 |
| 1.5 | 115 | 49 | 467 | 334 | 106 | 9.53 | 9 | 16 | 0.3 | 8 | 40.0 | 21.6 | 60 |
| Итого | 734 | 314 | 8801 | 6669 | 2589 | 28.0 | 39 | 123 | - | 109 | 160 | 93.8 | 262 |

Таблица 3

| No | $R_0$ | $R$ | $C_\Sigma$ | $C_{10}$ | $C_{max}$ | $C_s$ | $h$ | $g$ | $m$ | $I_{10}$ | $Kh_1$ | $Kh_2$ | $Kh_3$ |
|---|---|---|---|---|---|---|---|---|---|---|---|---|---|
| 1 | 2 | 3 | 4 | 5 | 6 | 7 | 8 | 9 | 10 | 11 | 12 | 13 | 14 |
| 2.1 | 61 | 37 | 465 | 347 | 97 | 12.6 | 12 | 16 | 0.75 | 12 | 34.8 | 21.6 | 49.6 |
| 2.2 | 132 | 78 | 614 | 252 | 74 | 8 | 12 | 16 | 0.24 | 16 | 34.0 | 24.8 | 46.2 |
| 2.3 | 154 | 28 | 385 | 212 | 47 | 13.7 | 11 | 16 | 0.41 | 14 | 26.0 | 19.6 | 33.0 |
| 2.4 | 164 | 72 | 366 | 137 | 40 | 5.08 | 8 | 9 | 0.15 | 6 | 14.3 | 19.1 | 29.9 |
| 2.5 | 103 | 42 | 213 | 129 | 33 | 5.1 | 7 | 9 | 0.24 | 5 | 14.7 | 14.6 | 24.3 |
| 2.6 | 89 | 49 | 318 | 174 | 30 | 6.5 | 9 | 16 | 0.37 | 8 | 20.5 | 17.8 | 27.0 |
| 2.7 | 27 | 12 | 99 | 88 | 25 | 8.25 | 6 | 9 | 0.86 | 6 | 18.3 | 9.9 | 19.8 |
| 2.8 | 29 | 22 | 98 | 76 | 23 | 4.5 | 5 | 4 | 0.11 | 5 | 10.8 | 9.9 | 16.5 |
| 2.9 | 198 | 142 | 789 | 165 | 19 | 5.6 | 11 | 16 | 0.20 | 12 | 17.8 | 28.1 | 19.0 |
| 2.10 | 38 | 12 | 49 | 47 | 18 | 4.1 | 3 | 4 | 0.06 | 1 | 10.5 | 7.0 | 12.0 |
| 2.11 | 36 | 18 | 113 | 93 | 14 | 6.3 | 6 | 9 | 0.35 | 5 | 13.0 | 10.6 | 13.9 |
| 2.12 | 13 | 8 | 47 | 47 | 13 | 5.9 | 4 | 4 | 0.09 | 2 | 10.5 | 6.8 | 11.2 |
| 2.13 | 83 | 21 | 93 | 70 | 12 | 4.4 | 6 | 4 | 0.27 | 2 | 9.5 | 9.6 | 11.6 |
| Итого | 1127 | 541 | 3649 | 1837 | 97 | 6.74 | 20 | 33 | - | 75 | 38 | 135 | 97 |

Таблица 4

| No | $R_0$ | $R$ | $C_\Sigma$ | $C_{10}$ | $C_{max}$ | $C_s$ | $h$ | $g$ | $m$ | $I_{10}$ | $Kh_1$ | $Kh_2$ | $Kh_3$ |
|---|---|---|---|---|---|---|---|---|---|---|---|---|---|
| 1 | 2 | 3 | 4 | 5 | 6 | 7 | 8 | 9 | 10 | 11 | 12 | 13 | 14 |
| 3.1 | 21 | 8 | 43 | 43 | 13 | 5.4 | 5 | 4 | 1.75 | 1 | 9.8 | 6.5 | 10.5 |
| 3.2 | 9 | 8 | 36 | 36 | 10 | 4.5 | 4 | 4 | 0.22 | 1 | 8.3 | 6.0 | 9.0 |
| 3.3 | 21 | 6 | 30 | 30 | 8 | 5.0 | 4 | 4 | 0.75 | 0 | 7.5 | 5.5 | 7.6 |
| 3.4 | 39 | 9 | 23 | 23 | 8 | 2.6 | 2 | 1 | 0.05 | 0 | 4.4 | 4.8 | 6.5 |
| 3.5 | 72 | 11 | 41 | 40 | 7 | 3.7 | 4 | 4 | 0.50 | 0 | 6.3 | 6.4 | 6.9 |
| 3.6 | 13 | 4 | 12 | 12 | 7 | 4.0 | 2 | 1 | 0.28 | 0 | 5.3 | 3.5 | 5.8 |
| 3.7 | 8 | 2 | 8 | 8 | 7 | 4.0 | 1 | 1 | 0.14 | 0 | 5.2 | 2.8 | 4.2 |
| 3.8 | 5 | 1 | 4 | 4 | 4 | 4.0 | 1 | 1 | 0.50 | 0 | 4.0 | 2.0 | 2.8 |
| 3.9 | 87 | 6 | 11 | 11 | 3 | 1.8 | 2 | 1 | 0.06 | 0 | 2.5 | 3.3 | 3.0 |
| 3.10 | 22 | 3 | 6 | 6 | 3 | 2.0 | 2 | 1 | 0.05 | 0 | 2.5 | 2.4 | 2.7 |
| 3.11 | 1 | 1 | 3 | 3 | 3 | 3.0 | 1 | 1 | 1.0 | 0 | 3.0 | 1.7 | 2.4 |
| 3.12 | 3 | 1 | 2 | 2 | 2 | 2.0 | 1 | 1 | 0.5 | 0 | 2.0 | 1.4 | 1.7 |
| Итого | 301 | 60 | 219 | 218 | 13 | 3.65 | 7 | 8 | - | 2 | 9 | 14.8 | 15 |

### 4. Индексы цитирования для авторских коллективов

Для авторских коллективов несложно построить аналогичные индивидуальным функции цитирования $C(r)$ с монотонно убывающим числом цитирований работ, получить все их значимые параметры $R_0 = \sum_{i=1}^{I} R_{0i}$, $R = \sum_{i=1}^{I} R_i$, $C_\Sigma = \sum_{i=1}^{I} C_{\Sigma i}$, $C_{max} = \max\{C_{max,i}\}$, $C_s = C_\Sigma / R$ и вычислить соответствующие индексы цитирования $C_{10}$, $I_{10}$, $h$-индекс, $g$-индекс, $Kh_1$, $Kh_2$, $Kh_3$. Здесь $I$ - число авторов в коллективе. Для сопоставляемых авторских коллективов с функциями цитирования 1.1-1.5, 2.1-2.13 и 3.1-3.12 они приведены в последних строках таблиц 2, 3, 4, а для объединенного авторского коллектива – в таблице 5. Функции цитирования авторских коллективов 1.1-1.5, 2.1-2.13 и 3.1-3.12 представлены на рис. 4 сплошными ломаными линиями 1, 2, 3, а для объединенного авторского коллектива штриховой ломаной линией 4. Индексы Дж.Э.Хирша отмечены прямыми треугольниками, индексы Л.Эгга – перевернутыми треугольниками, индексы $Kh_1$ – квадратами, индексы $Kh_2$ – темными кружками, индексы $Kh_3$ – светлыми кружками. Исходящие из начала координат тонкие сплошные линии соответствуют авторским коллективам 1.1-1.5, 2.1-2.13 и 3.1-3.12, штриховые линии – объединенному авторскому коллективу и имеют тот же смысл, что и в разделе 3. Следует отметить, что в разных системах цитирования для авторских коллективов используются, в основном, параметры $R_0$, $R$, $C_\Sigma$, $C_s$, а также средние значения числа работ и числа цитирований, приходящихся на одного автора $R_{0a} = R_0/I$, $R_a = R/I$, $C_a = C_\Sigma/I$, где $I$ - число авторов в коллективе. Из представленных на рис.4 сопоставлений авторского коллектива 1.1-1.5 (ломаная 1) и объединенного коллектива (ломаная 4) видно, что функцию цитирования для коллектива, включающего авторов с широким диапазоном числа цитируемых работ и числа цитирований, определяют данные небольшого числа (здесь 5 из 30) наиболее цитируемых авторов. Данные «средне»- и «малоцитируемых» авторов составляют лишь «хвост» объединенной функции цитирования и практически не влияют на $h -$ и $g -$ индексы объединенного коллектива, из-за чего последние не могут служить объективной характеристикой цитируемости таких авторских коллективов. Это принципиально искажает роль средне- и малоцитируемых авторов (в основном средневозрастных и молодых ученых) и реальный потенциал научных коллективов. В нашей стране отсутствие выехавших за рубеж за последние двадцать с лишним лет и ушедших в коммерческие структуры поколений отрицательно сказывается до сих пор. Индексы $Kh_1 - Kh_3$ позволяют учесть влияние данных «средне»- и «малоцитируемых» авторов, однако, это влияние проявляется по разному. Увеличение общего числа работ (за счет работ с заметно меньшими числами их цитирования) снижает значение $C_s$ и, соответственно индекс $Kh_1$ у объединенного коллектива. Индекс $Kh_2$ (т.е. $\sqrt{C_\Sigma}$) является лишь сопоставимой с $h -$ индексом модификацией параметра $C_\Sigma$ и вряд ли подходит в качестве эффективной альтернативы индексам Дж.Э.Хирша и Л.Эгга. Из числа рассматриваемых здесь, индекс $Kh_3$, который также, как $h -$ и $g -$ индексы, не зависит от числа цитируемых работ, является наиболее объективным критерием ранжирования авторских коллективов, в том числе свободным от манипулирования параметром $R$. Возможно, наиболее приемлемым (компромиссным) вариантом было бы использование индекса $Kh = \max\{Kh_1, Kh_2, Kh_3\}$.

Таблица 5

| No | $R_0$ | $R$ | $C_\Sigma$ | $C_{10}$ | $C_{max}$ | $C_s$ | $h$ | g | m | $I_{10}$ | $Kh_1$ | $Kh_2$ | $Kh_3$ |
|---|---|---|---|---|---|---|---|---|---|---|---|---|---|
| 1 | 2 | 3 | 4 | 5 | 6 | 7 | 8 | 9 | 10 | 11 | 12 | 13 | 14 |
| 1-3 | 2162 | 915 | 12669 | 8724 | 2589 | 13.8 | 42 | 123 | - | 186 | 132 | 113 | 290 |

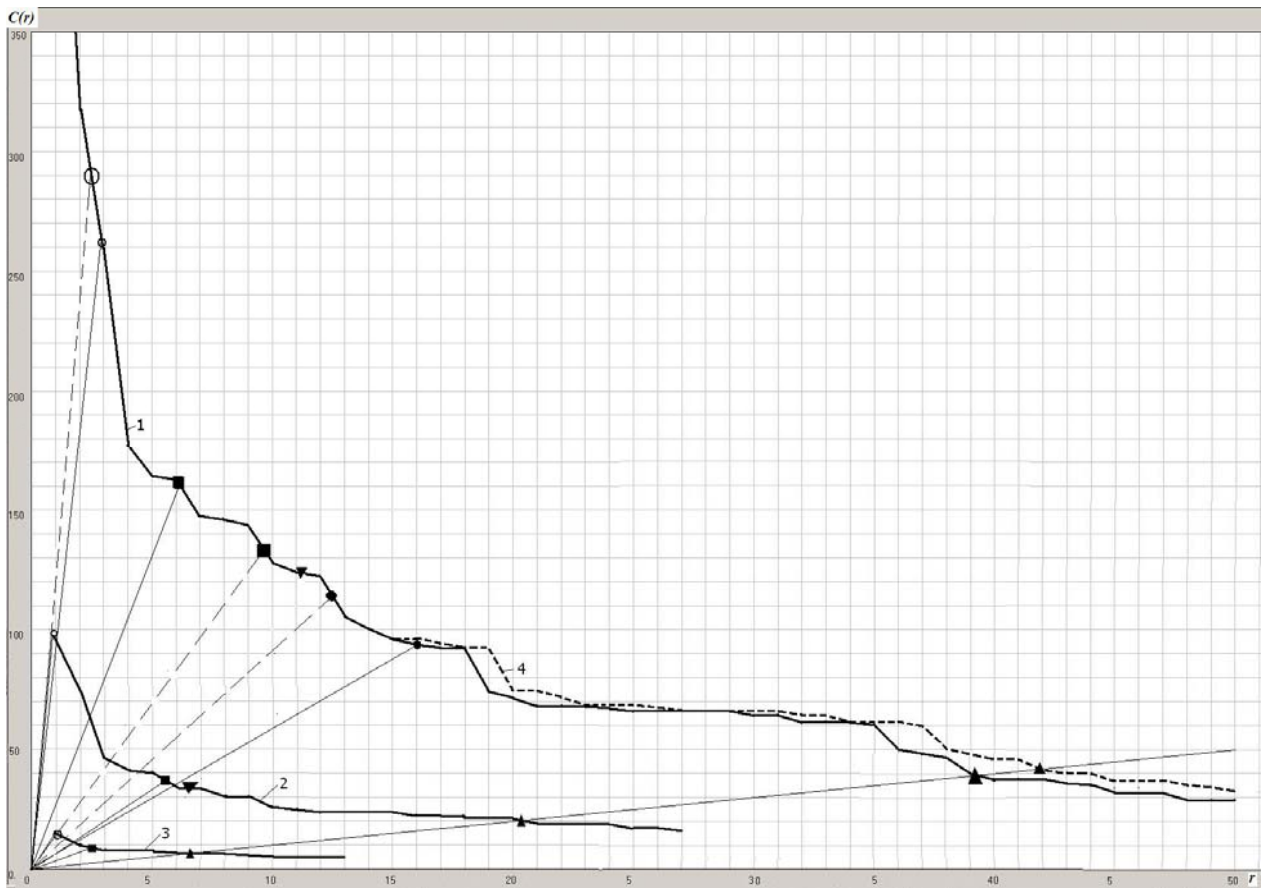

Рис.4. Индекс Дж.Э.Хирша – прямые треугольники; индекс Л.Эгга – перевернутые треугольники; индекс $Kh_1$ – квадраты; индекс $Kh_2$ – темные кружки; индекс $Kh_3$ – светлые кружки. Тонкие сплошные линии соответствуют авторским коллективам 1,2,3, штриховые линии – объединенный авторский коллектив.

**Заключение**

На основании приведенных данных и проведенного выше их анализа были установлены:
- для существенного повышения полноты и достоверности информации необходимость совершенствования автоматизированных систем цитирования научных работ путем максимального расширения возможностей авторских корректировок собственных данных с использованием доступных и удобных для авторов инструментов;
- необходимость введения (на уровнях редакционной обработки статей и работы систем цитирования) определенных ограничений по количеству и соответствию содержанию работы самоцитирований (не исключая их полностью, что принесло бы еще больший вред);
- необходимость ограничения возможностей других манипуляций с цитированиями (перекрестное цитирование, неоправданное расширение авторских коллективов и др.);
- необходимость полного исключения в системах цитирования учета импакт-факторов журналов (ранжирования по типу публикации), являющихся чисто издательскими параметрами и не влияющими на значимость научной публикации;
- необходимость исключения явной или косвенной языковой и иной возможной дискриминации в оценках научных публикаций (исключение из формируемых баз данных не англоязычных источников или игнорирование их поисковыми и обрабатывающими системами и др.);
- необходимость предоставления автору выбора устраивающей его системы цитирования во всех затрагивающих его интересы случаях, что неизбежно приведет к унификации систем цитирования и существенному упорядочиванию существующих в данной области проблем;
- необходимость совершенствования и унификации индексов цитирования в направлении большей их объективности, в том числе их сопоставлений только внутри каждой из областей знаний и возрастных групп.

**Цитируемая литература**